\documentclass[final] {aipproc}
\usepackage{sidecap,graphicx,boxedminipage,epsfig,wrapfig,floatflt,shadow}

\layoutstyle{8x11double}

\begin{document}

\title{ \bf{Interactive video tutorials for enhancing problem solving, reasoning, and meta-cognitive skills of introductory physics students}}

\author{Chandralekha Singh}{
  address={Department of Physics and Astronomy, University of Pittsburgh, Pittsburgh, PA, 15260, USA}
}

\begin{abstract}


We discuss the development of interactive video tutorial-based problems to help  
introductory physics students learn effective problem solving heuristics.  
The video tutorials present problem solving strategies using concrete examples in an interactive environment.  
They force students to follow a systematic approach to problem solving and students are 
required to solve sub-problems (research-guided multiple choice questions) to show their level of understanding at every stage of prob
lem solving.  
The tutorials are designed to provide scaffolding support at every stage of problem solving as needed
and help students view the problem solving process as an opportunity for knowledge and skill acquisition rather than a ``plug and chug
" chore. 
A focus on helping students learn first to analyse a problem qualitatively, and then to plan a solution in terms of the relevant 
physics principles, can be useful for developing their reasoning skills. The reflection stage of problem solving can help students
develop meta-cognitive skills because they must focus on what they have learned by solving the
problem and how it helps them extend and organize their knowledge.
Preliminary evaluations show that a majority of students who are unable to solve the tutorial problems without help can solve similar 
problems after working through the video tutorial. Further evaluation to assess the development of useful skills in tutorial learners 
vs. control group is underway.
\end{abstract}

\maketitle

\section{Introduction}

Problem solving can be defined as any purposeful activity where one is presented with a novel situation and  
devises and performs a sequence of steps to achieve a set goal.[1] Effective problem solving begins with a 
qualitative analysis of the problem, followed by planning, implementation, assessment, and reflection. 
As the complexity of a physics problem increases, it becomes increasingly important to employ a systematic approach.  
In order to help introductory physics students learn effective problem solving heuristics and develop useful skills,  
we are developing interactive video tutorial based problems. The tutorials force students to follow a systematic approach to problem
solving and provide appropriate guidance and feedback with a focus on helping them develop self-reliance.

Cognitive research indicates that the strategies used for effective acquisition, retention and retrieval of knowledge 
from memory have implications for helping students develop useful skills.[2] 
Students do not automatically develop useful skills by spending lots of time solving problems. 
There is evidence to suggest that the crucial difference between expert and novice problem solving lies 
in both the level and complexity of knowledge representation and rules.[1] 
Experts typically start with an initial plan which provides overall structure. 
Unlike novices who focus on surface features and jump into the implementation phase of problem solving immediately without thinking if
 a 
concept is applicable, experts concentrate on deep features and start with initial qualitative analysis and planning steps before reso
rting 
to the implementation issues.  Experts are relatively comfortable going between different representations of knowledge 
(e.g., verbal, visual, algebraic) and employ representations that make problem solving easier.[2] 
Also, since the typical ``chunk-size" for knowledge required for problem solving is larger for experts 
(they have lots of compiled knowledge),  they are able to reflect on the problem solving process while solving problems without 
experiencing cognitive overload. 
Because students' "knowledge chunks" are smaller than that of experts,  the limited capacity of short term memory makes the
cognitive load high during problem solving tasks, leaving few cognitive resources available for meta-cognition.
The abstract nature of the laws of physics and the chain of reasoning required to draw meaningful inferences makes these issues critic
al.
Appropriate scaffolding during problem solving can reduce cognitive load for students, and provide opportunities for metacognition. 

Video tutorials are being designed with these issues in mind so that they can help students learn problem solving, reasoning and 
meta-cognitive skills using concrete examples in an interactive environment.  The tutorials are designed to force students to analyse 
the problem
qualitatively and spend time deciding why certain principles of physics are appropriate. Consistent use of qualitative analysis
and planning tasks can help students develop reasoning skills. Also, helping students reflect upon the problem solving process at the 
end of
every problem and forcing them to think about what they learned by solving the problem and how it helps them restructure, extend and
organize their knowledge can help develop their meta-cognitive skills.

Students working on the tutorials first use a worksheet which divide the problems into five stages involved in problem solving. 
After attempting the problem on the worksheet to the best of their ability, students  access the same problem on the computer.  
Before inputting their numerical or symbolic solution on-line, students are required to solve several  
research-guided sub-problems to show their level of understanding at every stage of problem solving.  
The alternative choices in these multiple-choice questions elicit common difficulties students have with relevant concepts.[3]
Incorrect responses direct students to a short video (see the figure) while correct responses give them a choice of either advancing t
o  
the next sub-problem or watching videos to learn why the alternative choices are incorrect. 
While some videos are problem-specific, many focus on more general ideas, such as contact and non-contact forces or  
conditions for determining when a system is in equilibrium.  
Students have the option of watching additional videos which demonstrate and exemplify 
a particular problem solving stage such as how to perform qualitative analysis. 

Some other web-based tutorials have been developed, e.g., Cyber Tutor.[4]  The novel feature of our tutorials is the
incorporation of video, audio, and cursor movement to scaffold student learning. In addition, our sub-problems are typically more qual
itative
than Cyber Tutor and our alternative choices for the subproblems are guided by research in physics education.

\section{Details of Video Tutorials}

The main skeleton of interactive video tutorial based problems are being developed using Macromedia Flash.  
The videos focus on different topics in introductory physics. Most tutorials are suitable for students in any algebra or calculus-base
d
introductory physics course. Since the guidance and feedback is customized to students' needs, they 
can be used as aids to problem solving in homework and as a self-study tool by a diverse group of students.  
They can be helpful to students with different learning styles and  can help them view the problem solving process as an opportunity f
or knowledge and skill acquisition rather than a ``plug and chug" chore or guessing task. 
 
The development of the video tutorial based problems is guided by research. 
For the past few years, we have investigated if instruction which is based upon  
a field-tested cognitive apprenticeship model [5] of learning involving  
modeling, coaching, and scaffolding can help students learn effective problem solving heuristics and found positive outcome. 
In this approach, ``modeling" means that the instructor demonstrates and exemplifies the skills that students should learn. 
``Coaching" means providing students opportunity, guidance and practice so that they learn 
the skills necessary for good performance. 
``Scaffolding" means providing students with appropriate support and immediate feedback with a focus on weaning or fading the 
support so as to help them develop gradual self-reliance. 
This scaffolding process lies at the heart of Vygotsky's notion of stretching a student's ``zone of proximal development" 
and Piaget's idea of providing ``optimal mismatch".[6]
Using a similar instructional model, Schoenfeld has shown that in the context of mathematics, 
an explicit instruction in problem solving heuristics significantly improves students' skills.[6]

\begin{center}
\hspace*{-0.10in}
\epsfig{file=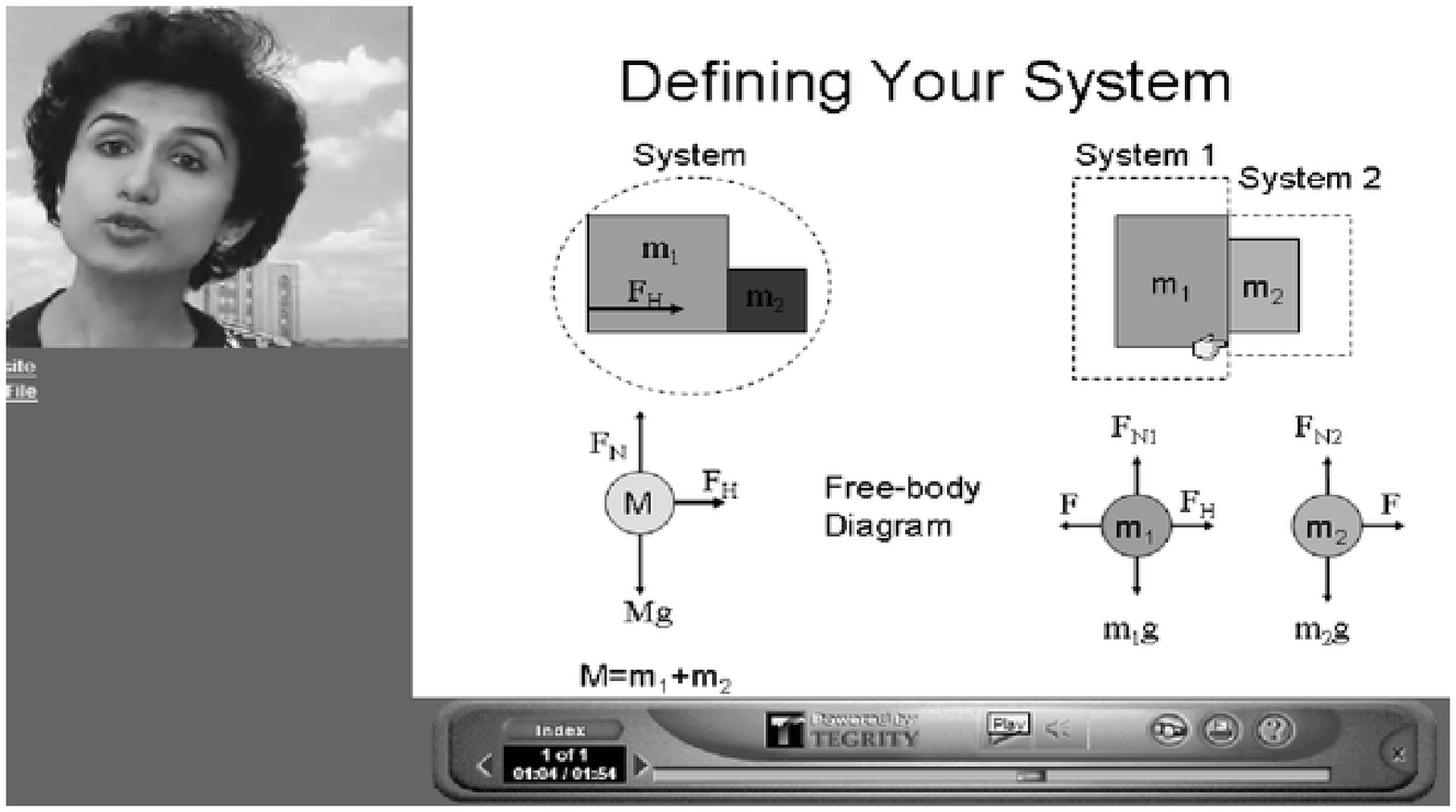,height=1.7in}
\end{center}

After the students have worked on the problem worksheet on their own, then worked on the tutorial and solved all the sub-problems upto
 the 
implementation-phase correctly, they are allowed to 
input their numerical or symbolic response with appropriate units. If they do not input the correct response in three attempts and it 
is 
apparent that their difficulty relates to mathematics rather than physics, e.g., how to solve 
two simultaneous equations, they may be directed to a video that helps with the relevant mathematics. Eventually, 
all students are given the option of watching the implementation phase of problem solving in detail.  
Then, students respond to a sub-problem which involves the reflection stage of problem solving.  
This sub-problem helps students reflect upon what they learned by solving the problem and how it can be applied in somewhat different 
contexts. 
If students have difficulty responding to this sub-problem, further videos will provide scaffolding support.  
Each video tutorial based problem is matched with another problem (paired problem) that uses  
similar physics principles. 
Students are assigned these problem pairs in homework to help them learn to decontextualize the problem solving 
approach and concepts learned from the video tutorial based problems. 
 
\section{
Preliminary Evaluation }

Below, we discuss the insights gained from task analysis and twelve student interviews 
for the design of subproblems and video tutorials in the context of a 
problem involving the non-equilibrium application of Newton's second law and Newton's third law. 
The problem statement is as follows:
Two adjacent boxes ($m_1=5 kg$, $m_2=2.5 kg$) are in contact on a horizontal frictionless table.
You apply a constant horizontal force $F_H=10 N$ to the larger block continuously.
Find the force $F$ exerted on the larger block by the smaller block.
 
We started with a cognitive task analysis which involves making a {\it fine-grained} 
flow chart of all the concepts that students must know 
{\it in the appropriate order} in which they must be invoked to accomplish the task (solve the problem). 
A careful task analysis 
is extremely helpful in identifying and addressing stumbling blocks or subtleties that students are likely to miss.
We then interviewed four students using think-aloud protocol 
about several problems involving equilibrium and non-equilibrium
applications of Newton's second law and Newton's third law (including the problem above). Based upon theoretical
task analysis and student interviews, we designed the preliminary version of the subproblems and video tutorials.
In eight later interviews using think aloud protocol, students were first given the worksheet and asked to solve the problem
to the best of their ability. Then, they were asked to attempt the subproblems and watch the videos as needed. 
To evaluate video tutorial's effectiveness,
they were then asked to solve two other similar problems requiring near- and far-transfer
on the worksheet without any help.

The above problem is apparently so difficult that none of the 12 interviewed students could solve it without help from video tutorials.
Many interviewed students, including those with A and B grades in their midterms on related concepts, had great difficulty 
figuring out whether this was an equilibrium or a non-equilibrium problem. Many believed that unless the problem
mentions it, they should take the net horizontal force to be zero on each object. 
A majority believed that the force exerted by the hand, $F_H$, should act on both boxes 
because the box with mass $m_2$ is in the way of the larger box that is being pushed.
Many had difficulty understanding why $m_2$ will exert a force on $m_1$. Some explicitly mentioned that the
reason they are confused is that the surface is frictionless here and that is the only force that will
push back on $m_1$ when they exert a force $F_H$. 
Newton's third law is extremely difficult for students and they had great difficulty realizing that the force exerted by
$m_1$ on $m_2$ is equal in magnitude to the force exerted on $m_2$ by $m_1$. Many felt that 
the larger object should exert a larger force on the smaller object than vice versa. Some justified their argument by
noting that $F_H$ is the force exerted on $m_2$ by $m_1$ and $F$ is the force exerted on $m_1$ by $m_2$ and it makes sense that
the larger box $m_1$ is exerting a larger force $F_H$ on the smaller box $m_2$ (students often felt that $F$ and $F_H$ are action-reaction
pairs in Newton's third law).

A majority could not make up their minds about what should be the system for which they should draw a free body diagram (FBD). 
Most did not understand that you can choose either $m_1$, $m_2$ or both of them together as a system depending upon your convenience
and based upon the choice of the system, the forces exerted by $m_1$ on $m_2$ and vice versa become internal forces that cancel in
pairs or external forces that should be taken into account. Several started drawing FBDs (with forces around a dot)
but had difficulty committing to the system that the diagram was for. Several were not clear about the fact that FBDs
should only include forces exerted on the system of choice by objects in the environment and not the forces exerted by the system
on objects in the environment. This kind of confusion often led students who noted they were drawing FBD of $m_1+m_2$
to include the force exerted by $m_2$ on $m_1$; an internal force for this system.
Although the students were recruited from different sections with different instructors, 
almost all of them had great difficulty with non-equilibrium applications of Newton's second law and Newton's third law.
Some students could only recall $F_{net}=0$ while others did not know what to do about $``a"$ after writing down $F_{net}=ma$.
A majority of students by themselves could not translate their FDBs into algebraic representation.
While some explicitly noted that the two boxes will have different accelerations since they have different masses, others said that
the net force on them should be the same since they move together. Some students who drew the force
$F$ exerted by $m_2$ on $m_1$ in the FBD in addition to $F_H$ wrote $F_H=m_1 a$. When asked why they did not include $F$ in the
equation, some were perplexed while others wrote a separate equation with $-F=m_1 a$.

After solving the subproblems and watching the videos as needed, all eight students opted to watch the video
related to the implementation phase of problem solving in detail (students knew that they would have to solve other problems
involving similar concepts without any help from videos). All students correctly solved the reflection subproblem
that dealt with the hand applying a force $F_H=10 N$ in the opposite direction on the smaller box.
The first problem that they had to solve without any help (after the tutorial) related to three 
boxes in contact on a frictionless surface and one pushed by a horizontal force. Students had to find the magnitude of the mutual forc
e between
the boxes. All students were able to draw the correct 
free body diagrams for the three boxes and write down Newton's second law equations for the non-equilibrium situation. 
Due to the time constraints during the interviews, once students drew the free body diagrams and wrote down 
Newton's second law equations correctly, they only had to explain verbally how they would solve them to find the required quantities. 
The only difficulty some students (typically those with lower course grade) had was whether the force exerted by
box 1 on 2 is equal in magnitude to the force exerted by box 2 on 1.  
This finding motivated us to include additional subproblems  that specifically deal with Newton's third law. The next problem they had
to solve without help related to three boxes on a frictionless horizontal surface and connected to each other via massless ropes and o
ne being  
pulled horizontally. Students had to find the magnitudes of tension force in the ropes. 
Several students explicitly noted that the reason this problem is similar to the one they had just solved 
is because the boxes exert forces on each other though the rope and they set up the problem correctly. However, two students 
got somewhat confused about the directions of tension forces while drawing the free body diagrams individually. 
Think-aloud protocol shows that while students were doing the reflection sub-problem and pair-problems, they were explicitly
trying to make connection with the tutorial problem and articulating, for example, how Newton's second law or third should apply to
that problem or how contact force of hand should not act on the object that the hand is not touching.

\section{Summary and Future Plans}

Teaching students to plan, reflect and draw qualitative inferences from quantitative problem solving is important for developing their
 skills. 
The need-based and self-paced nature of the research-guided video tutorials along with rewinding and stopping ability makes  
them suited for all students in introductory algebra- and calculus-based courses. Preliminary evaluations show that a majority of 
students who could not solve the tutorial problem on their own were able to solve similar problems on their own after working through 
video 
tutorials. Future evaluations will compare the development of skills in video tutorial learners with a control group that learned the 
same material for the same period of time
using other means. Eventually, we also plan to pursue quesions such as: (a) Can students still do original or similar problems after a
 couple 
of weeks? (b) When confronted with a problem they cannot do, do students consider the videos as a useful resource?

\bibliographystyle{aipproc}
{}
\end{document}